## Chapter XIII

# Communities of Practice: Going Virtual

### Chris Kimble, Paul Hildreth, and Peter Wright
### University of York, UK

*With the current trends towards downsizing, outsourcing and globalisation, modern organisations are reducing the numbers of people they employ. In addition, organisations now have to cope with the increasing internationalisation of business forcing collaboration and knowledge sharing across time and distance simultaneously. There is a need for new ways of thinking about how knowledge is shared in distributed groups. In this paper we explore a relatively new approach to knowledge sharing using Lave and Wenger's (1991) theory of Communities of Practice (CoPs). We investigate whether CoPs might translate to a geographically distributed international environment through a case study that explores the functioning of a CoP across national boundaries.*

## INTRODUCTION

As globalisation affects business, many organisations have taken steps to outsource and downsize in an effort to remain competitive (Davenport and Prusak 1998; O'Dell and Jackson Grayson 1997). Both downsizing and outsourcing mean not only a reduction in staffing levels with a corresponding loss of knowledge, but also an increase in the need to share knowledge across a distributed environment. . Increasingly, this knowledge is seen as central to the success of organisations and an asset that must be managed.

Several views of knowledge have been explored in Knowledge Management (KM) literature, most of them in the form of opposites. For example tacit/explicit (Nonaka 1991; Nonaka and Konno 1998); tacit/focal (Sveiby[1] Conklin[2]); know-what/know-how (Seely Brown and Duguid 1998), cognitivist/constructionist (von Krogh 1998) and work in practice and domain knowledge (Sachs 1995). Leonard and Sensiper (1998) however prefer to view knowledge as a continuum rather than a pair of opposites. They regard the two extremes as being tacit knowledge that is unconscious and held within people's minds, and totally explicit which is codified and structured.  They observe that most knowledge will reside somewhere between the extremes.

In this paper, we will differentiate between 'hard' and 'soft' knowledge.  Like Leonard and Sensiper we do not view them as mutually exclusive opposites, however neither do we view them as a continuum.  We view hard and soft knowledge as being two parts of a duality.  That is all knowledge is to some degree both hard *and* soft.

Harder aspects of knowledge are those aspects that are more formalised and that can be structured, articulated and thus 'captured'.  Soft aspects of knowledge on the other hand are the more subtle, implicit and not so easily articulated. For example a person's experience which allows them to make new inferences when presented with a new situation.

There are at least two forms of soft knowledge that can be identified.  The first is socially constructed knowledge.  In anthropology, socio-psychological and sociological work knowledge tends to seen as a product of social activity.  For example, Bruner (1990) argues that we should move away from the notion of the individual merely as a processor of information.  Instead we should move the emphasis to meaning and how this is negotiated in a community, as individuals cannot exist independently of their culture.  Similarly, Hutchins (1995) in developing his theory of Distributed Cognition also notes that looking for knowledge structures inside the individual fails to recognise that the social cultural environment always affects human cognition.  The second form of soft knowledge might be termed internalised domain knowledge.  Examples of this kind of soft knowledge might be skill, expertise and experience which has become second nature.  Winograd and Flores (1986) describe such knowledge as 'lost in the unfathomable depths of obviousness'.  It is this form of soft knowledge that is of prime interest in this paper.

## MANAGING HARD AND SOFT KNOWLEDGE

Much of the KM literature still takes the view that knowledge is 'hard' and concentrates on the capture-codify-store cycle.  In this sense, KM does not seem to have moved on from what was previously termed Information Management.  For example the view of knowledge as being 'hard' that is codifiable has led to attempts to extract knowledge from one group of 'experts' so that it can be used by another less skilled group.  The results of such systems however have been disappointing (Roschelle 1996; Schmidt 1997; Davenport and Prusak 1998).  Despite its evident problems, the management of 'hard' knowledge is now well established and there are many tools and frameworks available for this form of KM.

The soft knowledge embedded in the day-to-day working practices of communities is however much less amenable to a capture-codify-store approach.  Sierhuis and Clancey (1997) are explicit: they state that knowledge cannot be separated from the people and the situation.  Wenger (1998) too, stresses that 'information stored in explicit ways' is only a small part of the picture and that knowing is primarily something which comes about by participation in communities.  Clearly, we need to understand soft knowledge, how it is created sustained, and shared.  The recent work of Lave and Wenger (1991) provides a starting point.  The unit of analysis they suggest for this is a Community of Practice.

# COMMUNITIES OF PRACTICE

Lave and Wenger (1991) first introduced the concept of a Community of Practice in 1991.  Although the examples given (non-drinking alcoholics, Goa tailors, quartermasters, butchers and Yucatan midwives) were concerned with apprenticeship the central concept is not restricted to this form of learning.

Lave and Wenger (1991) described a Community of Practice as "… a set of relations among persons, activity and world, over time and in relation with other tangential and overlapping CoPs" (p98).  In such a community, a newcomer learns from old-timers by being allowed to participate in certain tasks that relate to the practice of the community.  Over time the newcomer moves from peripheral to full participation.  They regard a Community of Practice as "an intrinsic condition for the existence of knowledge" (p98).  They do not see the learning that takes place in such communities as narrow situated learning where instances of practice are simply replicated but as Legitimate Peripheral Participation (LPP).

LPP is both complex and composite.  Lave and Wenger (1991) explain that the three aspects, legitimation, peripherality and participation are indispensable in defining each other: they can not be considered in isolation.  Legitimation and participation together define the characteristic ways of belonging to a community whereas peripherality and participation are concerned with location and identity in the social world.  LPP is not merely learning situated in practice but learning as an integral part of practice.  Although the composite character of LPP is important, it is useful as an analytical convenience to consider the three components and their relationships separately.

Legitimation is the aspect that is concerned with power and authority relations in the community.  In the studies, legitimation is not necessarily formal.  For quartermasters, tailors and butchers there is some degree of formal legitimacy from hierarchy and rank but for the midwives and alcoholics legitimacy is more informal.

Peripherality is not a physical concept as in central and peripheral nor is it a simple measure of the amount of knowledge that has been acquired.  The terms peripheral and full participation are used to denote the degree of engagement with and participation in the community.  Lave and Wenger (1991) note that peripherality "must be connected to issues of legitimacy of the social organisation and control over resources if it is to gain its full analytical potential" (p37).

For Lave and Wenger (1991), participation provides the key to understanding CoPs.  A CoP does not necessarily imply co-presence, socially visible boundaries or a well-defined or identifiable group.  However, it does imply participation in an activity where participants have a common understanding about what it is and what it means for their lives and community.  The community and the degree of participation in it are inseparable from the practice.

## Extensions to the Community of Practice Concept

Many companies are increasingly turning to international teams (Castells 1996; Lipnack and Stamps 1997; West, Garrod and Carletta 1997) to improve their effectiveness when operating in the modern distributed international environment.  Teams are regarded as an effective and flexible means of bringing both skills and expertise to specific tasks and problems.  Partly as a result of this the concept of a CoP has been extended from Lave and Wenger's (1991) model to include a wider range of definitions (Stewart 1996[3]; Orr 1990; Seely Brown and Duguid 1991, 1996). Although the theme of learning was a prime driver for the concept of a CoP in its initial form, it is the extended concept that is of interest in this paper.

There have been several attempts to define CoPs in the commercial environment and some attempts by consultancies to formalise them. Seely Brown and Solomon Grey[4] offered:

> 'At the simplest level, they are a small group of people … who've worked together over a period of time. Not a team not a task force not necessarily an authorised or identified group … They are peers in the execution of "real work". What holds them together is a common sense of purpose and a real need to know what each other knows'

Seely Brown and Duguid (1991) applied Lave and Wenger's (1991) ideas to an ethnographic study previously undertaken by Orr (1990). In his work, Orr studied a group of photocopier repair technicians from the perspective of their collective memory. His explanation of how the technicians repaired the photocopiers was based on their ability to share soft knowledge in a CoP by the telling of 'war stories'. When a technician could not complete a particularly difficult repair by simply following the manual, he called his supervisor and the two worked together until the problem was solved. They did this by telling 'war stories' about similar problems they had encountered.

The process of story telling enabled them to exchange their soft knowledge and arrive at a solution to the problem. Over time, this solution was passed around other technicians and became part of the community's stock of knowledge. They had not only solved a problem but had also created new knowledge and contributed to the development of the community. War stories serve to legitimate a newcomer as they move from peripheral to fuller participation. The stories they tell and the stories in which they feature are used to assess members' competencies.

We can discern three methods of soft knowledge construction in such communities. Firstly there is the gathering of domain knowledge (for example how to solve a particularly tricky problem). Secondly, the construction of knowledge of work practices specific to the community (for example knowledge of the idiosyncrasies of an individual machine and how they are catered for). Finally, there is the knowledge that the community constructs about the competencies of its members (for example through the appraisal of their war stories).

These three methods could be regarded as being the 'soft' equivalent of the capture-codify-store approach of hard knowledge management. In the CoP if a problem had to be solved the members would gather the domain knowledge by interaction and working together to solve the problem. On the other hand, hard knowledge would be gathered a lot more easily because it is of the form that can be expressed and articulated so it could be transmitted a lot more easily. It can then be codified – for example into a database or an 'expert' system. The soft knowledge in the CoP however is not codified as such but it may become embedded in the practices of the community. Finally, hard knowledge is stored – for example in databases, books or reports from where it can be retrieved easily. Soft knowledge can also become stored in the community – in the relationships between the members as the members get to know each other and develop confidence in each other.

Communities of Practice are central to the maintenance of soft knowledge but all the studies in the literature (for example, Seely Brown and Duguid 1991; Lave and Wenger 1991) describe co-located communities. The internationalisation of business means that many organisations now function in a distributed international environment. This raises the question: can CoPs continue to operate in such an environment i.e. can a CoP be virtual? For example, could war stories be exchanged over the Internet? Similarly, as LPP is central to Lave and Wenger's (1991) notion of a CoP, a major issue is how LPP would translate to a geographically distributed environment. Learning undertaken with LPP is situated, as is some of the knowledge created during problem solving. The reason for the situatedness will have some bearing on how easy it is for a CoP to move into the geographically distributed environment. If co-location is necessary simply because members share resources such as a document then it should translate to the distributed environment relatively easily. However, if the learning is situated because the face-to-face element is essential for learning how the job is done then the distribution will have more impact.

Moving to a virtual environment also raises the question of whether it will be more difficult to gain legitimacy in such a community but perhaps the most difficult area will be the facilitation of participation. Participation is central to the evolution of a community. It is essential for the creation of the relationships that help to build the trust and identity that define a community. We will now investigate how CoPs translate to a geographically distributed international environment through a case study that explores the functioning of a CoP across national boundaries.

# THE CASE STUDY

The case study is of the operation of the IT support management team in the research arm of a major international company. The CoP consisted of two core groups: UKIT in the UK and USIT in the US. One core consisted of four co-located members in the UK: Bill (the UKIT manager); Steve (manager of the Informatics team); David (manager of the network team) and Michael (manager of the PC support team). They had equivalents in the other core USIT in the USA. There was one member of the CoP in Japan. The CoP as a whole adopted the unofficial title "IITMan" (International IT Management Team) as their group identity.

The study concentrated on the UK core, which was identified as having a number of features of a CoP:

- A sense of common purpose;
- the official group evolved from a need which is driven by the members themselves;
- a strong feeling of identity;
- having its own terminology (group specific acronyms and nicknames).

Previous work (Hildreth, Wright and Kimble 1999) has shown that distributed CoPs appear to evolve in a three-stage process:

1. The distributed CoP evolves from an initial informal contact between its members or from an official grouping. It develops into a CoP because of the way the members interact and work together.
2. A co-located CoP may develop links with individuals in other locations who are doing similar work. These people may also be members of other CoPs.

3. The developing CoP may then link with a similar group possibly in another country.

This process is evident in the study where the UK core has developed links with the US core to the point where the members of both consider themselves a Community of Practice.

In order to explore the interaction, communication and collaboration within this CoP a participant observation approach was used to get first hand observations. As part of a long term study, a week was spent with the UK members observing their day to day work and the interactions between the CoP members.

The data from the observations was analysed by creating five models to provide different views of the data (see Beyer and Holtzblatt 1997 for further details). The creation of an Affinity (Beyer and Holtzblatt 1997) allowed the identification of themes and categories and the relationships between them. The analysis of the data yielded several interesting insights into the workings of a distributed CoP.

## Three days in the life of an international distributed Community of Practice.

*Monday*

Steve returns to his cubicle and updates his palmtop computer from his PC so that his MS Exchange files are synchronised. His next job is to work on the planning document. He has lists of objectives from the other members of the UK management team and he wants to merge them into a common form to feed into the planning document. He scans the documents, and spends 20 minutes correcting errors from the Optical Character Recognition (OCR). David arrives and informs Steve that Bill and Michael have not arrived yet for the next meeting. David and Steve discuss briefly what Steve is doing with David's objectives list, what David agrees with, and where he would prefer it be handled differently. Their conversation then moves onto the forthcoming e-meeting with their American peers, and how the planning document will help them. David suggests using the same headings as a similar document produced by the American group. They compare the two documents to see how they can adapt theirs so that there is consistency. After this informal interruption, David and Steve go to the meeting with Bill and Michael. The planning document is the topic of this meeting and Steve hands out the updated version on which he has been working. Bill wants to discuss the latest version that Steve has prepared and move from that to how they can present it to their American colleagues. They take the document as the starting point and discuss how to prioritise. From there, the discussion moves to what the drivers for the projects are; then to whether the emphasis should be on development or consolidation, and then to the Year 2000 problem, which is a major element of the planning document. The planning document again becomes the focus of the discussion and this cycle continues - issues arise and are discussed; problems are flagged up, and discussions continue around the document. They talk round the problems and arrive at solutions or plans of action, always bringing their attention back to the document: the focal point of the meeting. Michael is paged. He is due at an e-meeting with the overall manager of IITMan. He leaves. David, Bill and Steve remain chatting informally for a few minutes and then adjourn intending to continue the meeting the following morning.

Steve returns to his cubicle. With only fifteen minutes to go before leaving for home, Steve synchronises his lap top with his PC. He then packs up to go home where he spends more time melding the remaining objectives and documents into the single planning document.

### Tuesday

The day starts with a continuation of yesterday's meeting. Bill summarises what was said in order that they do not go over it again. Steve explains that he has done some more melding and got the raw data in. He explains the structure that he is aiming for which will make it easier to use in the meeting with the American group. They continue the discussion about priorities and the Year 2000 problem and then work their way down the document looking for immovables and externally imposed deadlines, making notes on their individual copies of the document. As they work their way through the document, it fires up discussion about different issues, technical problems and issues of timing. The aim of the document at this stage is to be able to use it in the forthcoming e-meeting with the American colleagues so that they can show what they have done in trying to identify areas for collaboration. At the end of the morning, they break for lunch. Steve will continue to work on the document taking into account what was discussed in the meeting.

### Wednesday

08.45. Bill arrives for the regular weekly meeting scheduled for nine o'clock. He and Steve chat about the planning document. Steve carried on working at home last night and has the document as up to date as possible for this evening's e-meeting. He has managed to keep it fairly consistent with a previous American document. Bill is pleased with the result and feels that it is starting to take shape, making a suggestion about the addition of a further column so it could be used as a communication tool. Steve has also added an extra section, which meets with Bill's approval. He feels that it is already better than what they did the previous year.

The meeting gets under way. Bill recaps on the document, and what came out of it. Before they get too far into discussing the document Steve raises a problem – Stewart in his team will be away on holiday so they might have problems with the Exchange server. How can they support it while Stewart is away? They conclude that although they will not be doing any development, they will be able to work out how to support it between them. If they are stuck, they can always call on their American colleagues for advice. As a matter of course, Stewart keeps the American people informed. While he is away, the UK people can keep an eye on it, and the American people can keep a more distant eye on it.

The discussion moves on to the planning document. Bill now feels that it will make a good communication tool but will not *yet* be adequate as an internal project tool. Michael points out that the document has now lost detail of priority and who will do what. Steve replies that it has not been lost – there are sections for that but he has just not had time to populate that area of the document. David asks what the significance of the asterisks is. Steve replies that they are simply mistakes in the OCR reading and he still has them to correct. They take the most pressing issue from the document - the Year 2000 problem preparations. They had started discussing this yesterday and carry on discussing the preparations in detail moving on to strategy.

David outlines the plans he has made for tackling the problem. One of his ideas is to create a dummy infrastructure for testing. They could also collaborate with their American colleagues on this. The discussion moves onto planning and timing. Referring to the planning document Bill suggests a solution to how they might tackle the planning and timing issues. They continue their discussion bringing it back to the document and trying a different approach with the document. They finally decide to keep the Year 2000 problem as the single immovable and plan everything else around it. They move on to comparing their planning document with the earlier document from America. The meeting ends and Steve goes again to adjust the structure of the document that will be used in the evening's e-meeting. They will emphasise that it is in a first draft stage.

4.00pm. The e-meeting is due to start. Bill is still setting up the equipment and, as he does so, is chatting with the three people in California (Dan, Linda, Jim). Steve helps set up NetMeeting. Michael helps set up the connections and checks with their USIT colleagues in LA that they can see what is on the smartboard[5]. The equipment is almost ready. Steve is working on NetMeeting and Michael is using a remote keyboard with the smartboard.

Now we get a picture of them on the smartboard. It is a small picture - not bad but not wonderful. We see three people in the American meeting room. They wave. They decide to move straight on to the planning document. Bill gives a bit of background, but Dan interrupts almost immediately as he can see the scope for joint projects. David starts going through the document, highlighting the parts which are particularly interesting for him. As he goes through the items, they realise there is a problem with encryption and US export regulations. Attention is turned to discussing the problem; they find that LA can solve it. They continue working through the document and come to the virus section. This triggers a debate. They discuss the virus protection; decide to collaborate and work towards a common solution. Michael hand writes this as an action item on the smartboard. They move on to the next item, which is messaging. This is something where they are sending out different signals, but they recognise it as an area where they would benefit from working together. Michael writes this as an action item on the smartboard. They move to the next item, but something Dan says moves LA off on a tangent. Dan raises another issue - a tool has been launched by the organisation that might be able to help them. The UK people had not heard of this, so he suggests that they go and look at some presentation slides that are on the Intranet. He feels it might be an area where they could usefully collaborate. Michael notes it as an action item. Bill is aware of the pressure of time and pushes the discussion on. They move back to the planning document and Michael starts going through his section. He reaches the part regarding the organisation's 'SpeedPush' technology. This triggers questions from LA. Michael goes through the rest of his section and then hands over to Bill who starts with the 'audit' section. As they work through, they find other areas where they can collaborate on joint action. When Bill has finished, Steve takes over. He reaches a technical section on 16 bit obsolescence. This sparks a discussion, which leads to LA deciding to do some joint work with people from UKIT so that they can leverage from them. Michael again notes the action items. The UK team seems to be six months ahead and they have the opportunity to work together to provide an IIT strategy. The next item again sparks a debate, with several people trying to get in to say something.  LA has several concerns, but Michael points out that they are mistaken in their understanding of how the technology under discussion works. Steve moves down through his section of the document to multimedia publishing tools and raises a point that has been causing a problem.  Through discussion of the topic, they find that LA has already has a solution to the problem.  One more issue is flagged up which is causing the Steve's team a problem.  Dan suggests that they have a local expert who could perhaps help.  They reach the end of the list and Dan asks if they can have a copy of the planning document so they can track the revisions. Michael tries to send it through NetMeeting – he explains the settings that LA have to change and he transfers it into one of their directories.  PA say they would prefer it to be e-mailed round.  They also decide to copy Chakaka (the Japanese member of the CoP) in on the e-mail.  They arrange the next meeting for three weeks later to start at 3.00pm UK time and 7.00am Pacific Time.  In the meantime, Dan would like to examine the document more closely in order to see if they can integrate both the UK and American planning documents into one document.  He and Bill agree to get together electronically before the main meeting to discuss this further.  LA sign off. Bill, Michael, Steve and David decide to book the room for an earlier time and set the equipment up in advance next time.  They lost twenty minutes of the meeting today which represents a high proportion when they practically only have a two hour window (from 4.00pm to 6.00pm) when they can expect to be in contact with their American peers.

# Findings

*Shared Documents*

The major point of interest to come out of the case study is the use of a document as a shared artefact for communicating and sharing soft knowledge within the community and across national and cultural boundaries.

This happened in three ways:

1) the process of creating the document allowed the members of the CoP to share their soft knowledge through interaction.  The creation of the document involved the UKIT core members in meeting and collaborating.  They discussed issues and applied their knowledge to the creation of the document. They felt that they learnt from this process – in a group interview they expressed the view that, even at an early stage, the planning was better than the previous year.  They had benefited from the *process*, rather from simply having a document.  This was demonstrated when they told of the previous year's experience: a document had been created as an end in itself, and then rarely, if ever, referred to again.

2) some of the individuals' soft knowledge may be embedded in the document, for example, David worked at the Year 2000 section specifying how it was going to be tackled.  As a result this particular section was a reflection of David's planning skills.

3) when the document was used to communicate with members of the US core it acted as a catalyst to interaction, flagging up problems and issues for discussion, to which the individuals then applied their soft knowledge.  The members of the group were able to apply their knowledge in the *process* of working with the document.

During the week spent with the UK team, the sharing of documents proved a central activity. One document was of particular interest: a planning document developed by the UK core of the group. This particular document was of interest because it was being created for one purpose, a planning aid, but it was also used for other purposes, for example as a communication tool, both with the US core, and with the vertical teams (PC Support, Infrastructure, Informatics). The shared document was not essential to their work but it was more important than the group had previously realised. The many roles of the document were interesting: in particular its role in the creation and representation of knowledge.

The planning document represented the application of the soft knowledge both of the management teams and of their vertical teams. Each manager had received input from their team in the form of an e-mail or a formatted document. The individual managers then merged the input and created a planning document of their own. These planning documents were then brought to a management team meeting and discussed. One of the team then merged them into a draft planning document. During the week of the study, this draft document was the focus of three management team meetings during which it went through two more iterations. It was also used by at least one management team member to communicate with his team and drive a meeting.

The document was the subject of informal ad hoc discussion. Because of this, the structure was deliberately altered to take into account the need to communicate with the members in the US core. At the end of the week, there was a telephone conference with some US members of the group[6]. The focus of this meeting was the planning document that had been tailored for the purpose. During the meeting the members worked their way through the document which acted as a catalyst (rather than a vehicle) for the group members to apply their knowledge: for reflection, for solving problems, for discussion of issues and for planning.

A striking aspect of the document was its stimulative quality. Not only did it stimulate discussion of issues and the solving of problems but it also acted as a catalyst for collaboration. The group used the document to highlight areas where they could collaborate on projects where they could leverage from each other and where they could get their teams to work together. After the week of the case study, the two cores collaborated on the production of their two planning documents to provide a coherent collaborative plan. The document has since reached the stage where it is ready for consumption by other groups but it is part of an ongoing process of review thus it is still central to the work of the group and has become a living, ongoing document.

*Relationships – confidence, trust and identity - the extra half mile*
The artefact in the form of the planning document was deliberately designed to communicate effectively with the peers in the US crossing both the physical and cultural boundaries. The UK core were able to do this partly by using an existing document produced by UKIT, so they knew how the US document was structured. However, they were also in a position to design the document for cross boundary communication because they had already developed strong relationships with their peers and felt that they knew them very well. This was a feeling which was expressed in a later interview. In the later interview the group also felt that this relationship was particularly important as regards trust, as it gave them confidence in what they were receiving from each other, be it advice, information, help or documents — hence the UK core could develop the document knowing their peers in the US could have confidence in the contents.

The relationships had been developed over time and in most cases were based on having met the peers in a face to face environment. A lot of the community's work is undertaken within the UK and US cores but all the members meet physically on a six-monthly basis. In between these meetings, they maintain communication via e-media such as e-mail, voice mail, telephone, video link and Microsoft NetMeeting. The group members felt that they developed relationships with their colleagues and that they were able to get a lot of work done during the face-to-face meetings, for example one of the group, in a later interview, felt they could get through as much in one face-to-face meeting as in several e-meetings . They also felt that during the periods of communication by e-media the momentum gradually slowed until a physical meeting picked it up again. It has already been indicated in the section' Extensions to the Community of Practice Concept' that the development of relationships is essential to a CoP and that participation is key to developing the relationships. It was also indicated that participation might be a difficult aspect to maintain in a distributed environment. The case of IITMan shows that they try to maintain their relationships by the use of a range of e-media but that they also need to meet to refresh the relationship. When they meet they often find that they get through more work than is otherwise possible.

There are some interesting implications of the importance of a face to face element in a distributed CoP. The members felt they got to know each other better than they could by e-media. Having a good personal relationship with the other members was regarded as essential. A strong personal relationship was felt to be essential to carry the community through the periods of e-media communication. The members gained a greater feeling of unity and common purpose through knowing each other. One of the respondents described it as '… you need the personal relationship if you are to go the extra half mile for someone'. It was also felt that a strong personal relationship helped with issues of identity – the community members felt that they *knew* who they were communicating with even if it was via e-mail. Because they felt that they knew their partners so well, they were able to create the artefact that they wanted to share with their peers more effectively. They also had confidence in what they were receiving from them be it information, solutions to problems or simply opinions.

An essential point which arises from the study is that they do not consider CoPs to be (initially at least) formally created. In a formal group, such as a project group team or virtual team, legitimation comes from the group's formal structure. In a CoP, legitimation comes from social relationships that develop. As members get to know each other, they are better able to judge the information they receive from their partners. This shows the human aspect of a CoP to be of major importance.

# CONCLUSIONS

The literature has shown no reason why in theory a CoP might not be able to exist in a distributed international environment. There are some aspects that might cause difficulty, for example, in the section 'Extensions to the Community of Practice Concept' the question was raised as to how LPP might translate to a distributed environment. The research outlined in this paper has shown that CoPs can be maintained in the distributed environment although the CoP in the case study was not *totally* distributed as it had co-located cores. The central problem that the case study highlighted was that of sharing soft knowledge in this environment.

In the section on Extensions to the Community of Practice Concept we saw that participation is central to the evolution of a community and that it is essential to the creation of the relationships that help to build trust. The case study of IITMan has supported this view, for the respondents in interviews stressed the importance of relationship development and how this is easier through regular face to face interaction where participation is easier. A strong relationship is important for the members of the community to go the extra half mile for each other. For all practical purposes, the relationship is made in a face to face meeting: this enables the relationship to develop quicker and to go further.

Li and Williams (1999) and Ishaya and Macaulay (1999) have also stressed the importance of face to face communication - even in the modern distributed environment with a wide range of communications media. The findings of the case study support this as they showed the continued importance of maintaining face-to-face contact - it sustains subsequent e-communication but needs re-charging at intervals. This re-charging of the relationship then contributes to further evolution of the CoP as the relationship is grown further. As members of the CoP have increased confidence and trust in each other, so they gain legitimacy in each other's eyes.

The most striking finding of the case study was the importance of a shared artefact to the community and the range of uses to which it was put. It worked as a catalyst for collaboration; it was the focus of meetings and discussions and thereby highlighted a range of issues and problems; it was used for planning and co-ordination of work and it was also used as a communication tool.

The importance of the shared artefact and the development of strong relationships through participation strongly support more recent work by Wenger (1998) in which he identifies participation as being the essential component of LPP. However, he emphasises that participation is only one part of a duality, the other part being reification. By reification, he means giving experience form by creating artefacts that make the experience more concrete, for example, artefacts, books and stories. Wenger (1998) emphasises that participation and reification are a duality and one can therefore not exist without the other – there needs to be the correct balance between the two. Changing the balance changes the possibilities of how an artefact can be used, so the balance must be right for the practice of the community. It is also important to note that it is not the artefact per se which is important but the *process* involved in its creation. This, too, is borne out by the case study, where the process of creating the document and working with it is what was most beneficial to the CoP. They were able to share knowledge by both participating in the process of creating it, and by participating in the discussions and collaborations which resulted from it.

Although Wenger (1998) has modified the original concept of a CoP (Lave and Wenger 1991) his examples are still in the co-located environment. The case study reported here has explored the CoP concept in a distributed environment. Therefore, although the primary aim of the planning document was to help with planning and co-ordination, it was also intended that it would work as a communication tool with the US core. To that end, it was designed to take into account that it would inevitably be crossing boundaries.

In the case study there were different types of boundary which the artefact had to cross – it passed between different groups (it was used to communicate with a vertical team) it crossed the boundary between the cores and it had to cross cultural and national boundaries. Star (1989) and Star and Griesemer (1989) developed the notion of Boundary Objects to explain co-ordination work between communities. Boundary Objects cross the boundaries between communities and retain their structure but are interpreted differently. Boundary Objects show that although knowledge may be embedded in artefacts it is not a simple matter of capturing the knowledge and passing it on or of taking soft knowledge and making it 'hard': some abstraction takes place and some of the soft knowledge is lost in the process. Artefacts still need to be interpreted, that is, some domain knowledge is still necessary to be able to use the artefact and differing degrees of soft knowledge will enable to user to make different inferences from the artefact. Wenger (1998) has also explored shared artefacts as Boundary Objects as they crossed boundaries between communities.

In the case study as well as being translated across different groups and communities, the artefact was also translated across different media. Input to the document came from members of the vertical teams and was incorporated into separate documents produced by three of the CoP members. These were then merged to form a first version, which was discussed and adapted. The artefact also passed through e-mail, was worked on in different systems and was printed out for discussion, that is, it was propagated across different states, albeit not totally without problems: there were occasional technical problems to be overcome and occasional misunderstandings.

Although the shared artefact does not *solve* the problem of soft knowledge sharing by taking soft knowledge and passing it on, it has been shown that it can be of real benefit, play a variety of roles and be a catalyst in the sharing of soft knowledge even in the distributed environment. The research of which this case study is a part has supported Wenger's (1998) notion of participation and reification by highlighting (a) the importance and use of shared artefacts and (b) the importance of developing strong relationships through participation. Importantly however it has moved CoPs forward by exploring their operation in the distributed international environment and demonstrating that Wenger's (1998) notion of the reification/ participation duality could well provide a route forward.

# ENDNOTES

*assets in a knowledge economy*.   Available: http:www.zilker.net/business/info/pubs/desom/body.htm [1997 March 27th]

3.      Stewart T. A. (1996): The Invisible Key to Success. *Fortune Online*.  Available: http://pathfnder.com/@@V3AagAUAZyqOEYKS/fortune/magazine/1996.960805/edg.html [October 4th 1996]

4.      Seely Brown J & Solomon Gray E (no date): The People are the Company. *Fast Company* [Online] Available: http://www.fastcompany.com/online/01/people.html [September 9th 1998]

5.      Smartboard: cross between a white board and a computer screen.  It can show what is on a computer screen, it can be written on and it can be linked with other smartboards so users in different locations can all see what is on the board.

6.      This meeting also used a one way video link and NetMeeting.